\documentclass{article}
\usepackage{verbatim}
\usepackage{amsmath, amssymb}  
\newcommand{\paae}{p^{a}_{\alpha_1}}
\newcommand{\paat}{p^{a}_{\alpha_2}}
\newcommand{\pabe}{p^{b}_{\beta_1}}
\newcommand{\pabt}{p^{b}_{\beta_2}}

\newcommand{\pbae}{p^{a}_{\alpha_1}}
\newcommand{\pbat}{p^{a}_{\alpha_2}}
\newcommand{\pbbe}{p^{b}_{\beta_1}}

\newcommand{\pbbeae}{p^{b|a}_{\beta_1\alpha_1}}
\newcommand{\pbbeat}{p^{b|a}_{\beta_1\alpha_2}}
\newcommand{\pbbtae}{p^{b|a}_{\beta_2\alpha_1}}
\newcommand{\pbbtat}{p^{b|a}_{\beta_2\alpha_2}}

\newcommand{\pabeat}{p^{a|b}_{\beta_1\alpha_2}}

\newcommand{\pbaebe}{p^{b|a}_{\alpha_1\beta_1}}

\newcommand{\paaebe}{p^{a|b}_{\alpha_1\beta_1}}
\newcommand{\paaebt}{p^{a|b}_{\alpha_1\beta_2}}
\newcommand{\paatbe}{p^{a|b}_{\alpha_2\beta_1}}
\newcommand{\paatbt}{p^{a|b}_{\alpha_2\beta_2}}

\begin{document}

\title{On consistency of the quantum-like representation algorithm }
\author{Peter Nyman\\International Center for Mathematical Modelling\\ in Physics and Cognitive Sciences \\ University of V\"axj\"o, S-35195, Sweden\\
              {\bf E-mail:}{ peter.nyman@vxu.se}}
\maketitle

\abstract{In this paper we continue to study so called
``inverse Born's rule problem'': to construct representation of 
probabilistic data of any origin by a complex probability amplitude which 
matches Born's rule.  The corresponding algorithm -- quantum-like 
representation algorithm (QLRA) was recently proposed by A. Khrennikov 
\cite{IP1}--\cite{IP5}.  Formally QLRA depends on the order of conditioning. For
 two observables $a$ and $b,$  $b\vert a$- and $a \vert b$ conditional probabilities
produce  two representations, say in  Hilbert spaces $H^{b\vert a}$ and $H^{a\vert b}.$
In this paper we prove that under  natural assumptions these two representations
are unitary equivalent. This result  proves consistency QLRA.}

\begin{flushleft}
{\bf Keywords}{ quantum-like  representation algorithm, inverse Born's rule problem, 
order of conditioning, unitary equivalence of representations.} 
\end{flushleft}


\section{Introduction}

During last 80 years tremendous efforts were put to clarify inter-relation between 
classical and quantum probabilities, see, e.g., von Neumann \cite{VN} for the 
first detailed presentation of this problem and see, e.g., Gudder \cite{GD1}--\cite{GD3}, 
Svozil \cite{SV1}, \cite{SV2},  Fine \cite{F}, Garola et al. \cite{GAR1}--\cite{GAR3},
Dvurecenskij and Pulmanova \cite{DV}, Ballentine \cite{BL},  
O.  N\'an\'asiov\'a et al \cite{NAN}, \cite{NAN1},  Allahverdyan et al
\cite{AL}  for modern studies.\footnote{The list of references is far 
from to be complete, see Khrennikov's monograph \cite{IP} for the detailed list of references.}
We remark that during the last 30 years the main interest was attracted by Bell's inequality, see,
e.g., for detailed presentation. 
However, the basic rule 
of QM is  Born's rule. Therefore the study of its origin  is not less
(and may be even more) important than investigations  on Bell's inequality. In this paper 
we continue to study so called ``inverse Born's rule problem'' as it was 
formulated by Khrennikov \cite{IP1} --\cite{IP5}: 

\medskip

{\bf IBP} (inverse Born problem): {\it To construct representation of 
probabilistic data by a complex probability amplitude which matches Born's 
rule.} 

\medskip

Solution of IBP provides a possibility to represent probabilistic data by 
``wave functions'' and operate with this data by using 
linear algebra (as we do in conventional QM). In a special case (for a pair of dichotomous observables) this problem was solved in 
\cite{IP1} --\cite{IP5} with the help of so called quantum-like 
representation algorithm -- QLRA. 
Formally the output of  QLRA depends on the order of conditioning. For
 two observables $a$ and $b,$  $b\vert a$- and $a \vert b$ conditional probabilities
produce  two representations, say in  Hilbert spaces $H^{b\vert a}$ and $H^{a\vert b}.$
In this paper we prove that under  natural assumptions these two representations
are unitary equivalent. This result proves consistency of QLRA.

\section{Inversion of Born's Rule}

We consider the simplest situation. There are given two
dichotomous observables of any origin: $a=\alpha_1, \alpha_2$ and $b=\beta_1,
\beta_2.$  We set $X_a=\{\alpha_1,
\alpha_2\}$ and $X_b=\{\beta_1, \beta_2\}$ -- ``spectra of
observables".

We assume that there is given  the matrix of transition
probabilities ${\bf P}^{b\vert a}= (p^{b\vert a}_{\beta \alpha}),$ where
$p^{b\vert a}_{\beta \alpha}\equiv P(b=\beta \vert a=\alpha)$ is the
probability to obtain the result $b=\beta$ under the condition
that the result $a=\alpha$ has been obtained. There are also given probabilities 
$p^a_\alpha \equiv P(a=\alpha),
\alpha\in X_a,$ and $p^b_\beta \equiv P(b=\beta), \beta \in X_b.$
Probabilistic  data $C=\{ p^a_{\alpha}, p^b_{\beta}\}$ are related
to some experimental context (in physics preparation procedure).

IBP is to represent this data by a probability amplitude
$\psi$ (in the simplest case it is complex valued) such that
Born's rule  holds for both observables:
\begin{equation}
\label{BR} p^b_{\beta}= \vert \langle \psi, e^{b\vert a}_\beta\rangle
\vert^2\,,\qquad p^a_{\alpha}= \vert \langle \psi, e^{b\vert a}_\alpha\rangle
\vert^2\,,
\end{equation}
where $\{e^{b\vert a}_\beta \}_{\beta \in X_b}$ and $\{e^{b\vert a}_\alpha\}_{\alpha
\in X_a}$ are orthonormal bases
for observables $b$ and $a,$ respectively (so the observables are diagonal in respective 
bases). 

In \cite{IP1} --\cite{IP5} the solution of IBP is given in the form of an algorithm which 
constructs a probability amplitude from data. Formally, the output of this algorithm   
depends on the order of conditioning. By starting with the matrix of transition probabilities ${\bf P}^{a\vert b},$ instead of 
${\bf P}^{b\vert a},$ we construct another probability amplitude $\psi^{a\vert b}$
(the amplitude in (\ref{BR}) should be denoted by   $\psi^{b\vert a})$ and other bases, 
$\{e^{a\vert b}_\beta\}_{\beta \in X_b}$ and $\{e^{a\vert b}_\alpha\}_{\alpha
\in X_a}.$ We shall see that under natural assumptions these two representations are unitary 
equivalent.

\section{QLRA}
\subsection{$H^{b \vert a}$-conditioning}

Suppose that the matrix of transition probabilities  
${\bf P}^{b\vert a}$ is given. In \cite{IP1} --\cite{IP5}  the following formula for
interference of probabilities (generalizing the classical formula of total probability)
 was derived:
$ p^b_{\beta} \;=\; \sum_\alpha p^a_{\alpha} p^{b\vert a}_{\beta
\alpha} + 2 \lambda_\beta \; \sqrt{\prod_\alpha p^a_{\alpha}
p^{b\vert a}_{\beta \alpha}},
$
where the ``coefficient of interference"
\begin{equation}
\label{GGTT}
\lambda_\beta \;=\; \frac{p^b_{\beta} - \sum_\alpha
p^a_{\alpha} p^{b\vert a}_{\beta \alpha}}{2 \sqrt{\prod_\alpha p^a_{\alpha}
p^{b\vert a}_{\beta \alpha}}}\, .
\end{equation}
 We shall  proceed under the conditions:

(1) ${\bf P}^{b\vert a}$ is doubly stochastic.

(2) Probabilistic data $C=\{ p^a_{\alpha}, p^b_{\beta}\}$ consist 
of strictly positive probabilities.

(3)  Coefficients of interference $\lambda_\beta, \beta \in
X_b,$ are bounded by one:
$
\vert \lambda_\beta\vert \leq 1\,.
$

Probabilistic data $C$ such that (3)  holds is called {\it
trigonometric}, because in this case we have the conventional
formula of trigonometric interference\footnote{This formula can be easily derived 
in the conventional QM formalism, see, e.g., \cite{IP}, by transition 
from the basis of eigenvectors for the $a$-observable to the basis of eigenvectors for the
$b$-observables. We recall that in QM observables are given by self-adjoint operators.   
However, we proceed in the opposite 
way. We would like to produce a complex probability amplitudes and operator representation of 
the observables by using this formula.}:
$ p^b_{\beta} \;=\; \sum_\alpha p^a_{\alpha} p^{b\vert a}_{\beta
\alpha} + 2 \cos\theta_\beta \; \sqrt{\prod_\alpha p^a_{\alpha}
p^{b\vert a}_{\beta \alpha}}\,,
$
where
\begin{equation}
\label{GGTT1}
\lambda_\beta=\cos \theta_\beta\,.
\end{equation}
By using the elementary formula:
$ D=A+B+2\sqrt{AB}\cos \theta=\vert \sqrt{A}+e^{i
\theta}\sqrt{B}|^2, $ for real numbers $A, B > 0, \theta\in [0,2
\pi],$ we can represent the probability $p^b_{\beta}$ as the
square of the complex amplitude (Born's rule):
$ p^b_{\beta}=\vert \psi^{b\vert a}_\beta \vert^2 \,.$
Here
\begin{equation}
\label{EX1} \psi^{b\vert a}_\beta \;=\; \sqrt{p^a_{\alpha_1}p^{b\vert a}_{\beta \alpha_1}}
+ e^{i \theta_\beta} \sqrt{p^a_{\alpha_2} p^{b\vert a}_{\beta \alpha_2}}\,, \quad
\beta \in X_b\,.
\end{equation}
The formula (\ref{EX1}) gives the quantum-like representation
algorithm --- QLRA. For any trigonometric probabilistic data $C$,
QLRA produces the complex amplitude $ \psi^{b \vert a}$  (the normalized vector in 
the two dimensional complex Hilbert space, say $H^{b\vert a}):$ 
\begin{equation}
\label{EXY}
\psi^{b \vert a}  =
\psi^{b\vert a}_{\beta_1} e^{b\vert a}_{\beta_1} + \psi^{b\vert a}_{\beta_2} e^{b\vert a}_{\beta_2},   
\end{equation}
where
$
\label{BasTE} e^{b\vert a}_{\beta_1} \;=\;\left( \begin{array}{l} 1 \\
0
\end{array}
\right )\,,\qquad e^{b\vert a}_{\beta_2} \;=\;\left( \begin{array}{l}   0 \\
1
\end{array}
\right )\,.
$
 To solve IBP completely, we would like to have Born's rule not
only for the $b$-variable, but also for the $a$-variable:
$p^a_{\alpha}=\vert \langle \psi^{b\vert a}, e^{b \vert a}_\alpha \rangle\vert^2 \;,
\alpha \in  X_a.$ 
Here the $a$-basis in the Hilbert space $H^{b\vert a}$  is given, see \cite{IP1} --\cite{IP5} 
for details, by 
$
 e^{b\vert a}_{\alpha_1} \;=\;\left( \begin{array}{l} \sqrt{p^{b\vert a}_{\beta_1 \alpha_1}} \\
\sqrt{p^{b\vert a}_{\beta_2 \alpha_1}}
\end{array}
\right )\,,\qquad e^{b\vert a}_{\alpha_2} \;=\;\left( \begin{array}{l}  \; \sqrt{p^{b\vert a}_{\beta_1 \alpha_2}} \\
- \sqrt{p^{b\vert a}_{\beta_2 \alpha_2}}
\end{array}
\right )\,.
$
It is orthonormal, since ${\bf P}^{b \vert a}$ is assumed to be doubly stochastic.
In this basis the amplitude $\psi^{b \vert a}$ is represented as 
\begin{equation}
\label{PSI}
\psi^{b\vert a}= \sqrt{p^a_{\alpha_1}} e^{b\vert a}_{\alpha_1} +
e^{i \theta_{\beta_1}} \sqrt{p^a_{\alpha_2}} e^{b\vert a}_{\alpha_2}
\end{equation}

We recall that in QM a pure state $\Psi$ is defined as an equivalent class with respect to 
multipliers of the form $c=e^{i\gamma}.$ We shall use similar terminology.  
Each complex amplitude $\psi^{b \vert a}$ produced by QLRA determines 
a {\it quantum-like state} (representing given probabilistic data) -- the equivalence
class $\Psi^{b \vert a}$ determined by the representative $\psi^{b \vert a}.$

\subsection{$H^{a \vert b}$-conditioning}
Here
\begin{equation}
\label{EX1A} \psi^{a\vert b}_\alpha \;=\; \sqrt{p^b_{\beta_1}p^{a\vert b}_{\alpha \beta_1}}
+ e^{i \theta_\alpha} \sqrt{p^b_{\beta_2} p^{a\vert b}_{\alpha \beta_2}}\,, \quad
\alpha \in X_a\,.
\end{equation}
 For any trigonometric probabilistic data $C$,
QLRA produces the complex amplitude $ \psi^{a \vert b}$ (the normalized vector in 
the two dimensional complex Hilbert space, say $H^{a\vert b}):$ 
\begin{equation}
\label{EXYA}
\psi^{a\vert b}  =
\psi^{a\vert b}_{\alpha_1} e^{a\vert b}_{\alpha_1} + 
\psi^{a\vert b}_{\alpha_2} e^{
a \vert b}_{\alpha_2},   
\end{equation}
where
$
\label{BasTET} e^{a\vert b}_{\alpha_1} \;=\;\left( \begin{array}{l} 1 \\
0
\end{array}
\right ),\, e^{a\vert b}_{\alpha_2} \;=\;\left( \begin{array}{l}   0 \\
1
\end{array}
\right )\,.
$
Here the $b$-basis in the Hilbert space  $H^{a\vert b}$  is given by 
$
 e^{a\vert b}_{\beta_1} \;=\;\left( \begin{array}{l} 
\sqrt{p^{a\vert b}_{\alpha_1 \beta_1}} \\
\sqrt{p^{a\vert b}_{\alpha_2 \beta_1}}
\end{array}
\right ),\, e^{a\vert b}_{\beta_2} \;=\;\left( \begin{array}{l}  \; 
\sqrt{p^{a\vert b}_{\alpha_1 \beta_2}} \\
- \sqrt{p^{b\vert a}_{\alpha_2 \beta_2}}
\end{array}
\right )\,.
$
In this basis the amplitude $\psi^{a\vert b}$ is represented as 
\begin{equation}
\label{PSIA}
\psi^{a\vert b }= \sqrt{p^b_{\beta_1}} e^{a\vert b}_{\beta_1} +
e^{i \theta_{\alpha_1}} \sqrt{p^b_{\beta_2}} e^{b\vert a}_{\beta_2}
\end{equation}

As in the case of $H^{b \vert a}$-representation,
the {\it quantum-like state} (representing given probabilistic data) is defined as
 the equivalence class $\Psi^{a \vert  b}$  with the representative $\psi^{a \vert a}.$

\section{Unitary equivalence of \\ $b \vert a$- and $a \vert b$-representations}

Thus, as we have seen by selecting to types of conditioning, we represented the 
probabilistic data $C=\{p_\alpha^a, p_\beta^b\}$ by two quantum-like states,
$\Psi^{b \vert a}$ and $\Psi^{a \vert b}.$ We are interested in consistency of these
representations. 

We remark that any linear operator $W: H^{b \vert a} \to H^{a \vert b}$ induces 
the map of equivalence classes of the 
unit spheres with respect to multipliers  $c= e^{i \gamma}.$ 
We define the unitary operator $U_{b \vert a}^{a \vert b}: H^{b \vert a} \to H^{a \vert b}$
by $U(e^{b \vert a}_\alpha)= e^{a \vert b}_\alpha, \alpha \in X_a.$ It induces the mentioned 
map of equivalent classes.

\medskip
\begin{flushleft}
{\bf Theorem.} {\it The operator $U_{b \vert a}^{a \vert b}$ maps 
$\Psi^{b \vert a}$ into $\Psi^{a \vert b}$ if and only if the following 
inter- relation of symmetry takes place for matrices of transition probabilities 
${\bf P}^{b \vert a}$ and ${\bf P}^{a \vert b}$:
\begin{equation}
\label{PTTT}
p_{\beta \alpha}=p_{\alpha \beta },
 \end{equation}
 for all $\alpha$ and $\beta$ from spectra of observables $a$ and $b.$}
{\bf Proof.} Take the representative of $\Psi^{b \vert a}$ given by (\ref{PSI}). 
Then\end{flushleft} 
\begin{equation}\label{UPSI}
U_{b \vert a}^{a \vert b} \psi^{b \vert a}= \sqrt{p^a_{\alpha_1}} e^{a\vert b}_{\alpha_1} +
e^{i \theta_{\beta_1}} \sqrt{p^a_{\alpha_2}} e^{a \vert b}_{\alpha_2}
\end{equation}
Our aim is to show that this vector is equivalent to the vector $\psi^{a \vert b}$ given by 
(\ref{EXYA}).
By using $H^{b \vert a}$ analogs of (\ref{GGTT}) and (\ref{GGTT1}) 
for the coefficients of interference and its $\cos$-expression we 
determine $\cos \theta_{\alpha_1}:$
\begin{align}\label{PSI7}
 \paae&={\pabe \paaebe}+{\pabt\paaebt}+2\cos\theta_{\alpha_1}\sqrt{\pabe \paaebe \pabt\paaebt}\\\nonumber &\Leftrightarrow\\\nonumber
  \cos \theta_{\alpha_1}&=\frac{\paae-{\pabe \paaebe}-{\pabt\paaebt}}{2\sqrt{\pabe \paaebe \pabt\paaebt}}.
\end{align}
We also calculate 
\begin{eqnarray}\label{Psipsi}
\psi^{a|b}_{\alpha_2}\overline{\psi^{a|b}_{\alpha_1}}
&=\pabe\sqrt{\paaebe  \paatbe}-\pabt\sqrt{\paatbt\paaebt}\\\nonumber
&-(\cos\theta_{\alpha_1}+i\sin\theta_{\alpha_1})\sqrt{\pabt\paatbt\pabe \paaebe}\\\nonumber
&+(\cos\theta_{\alpha_1}-i\sin\theta_{\alpha_1})\sqrt{\pabe \paatbe\pabt\paaebt},
\end{eqnarray}
where $\psi^{a|b}_{\alpha_2}=\sqrt{\pabe \paatbe}-e^{i \theta_{\alpha_1}}\sqrt{\pabt\paatbt}$ is given by (\ref{PSIA}).
We use that  $|\langle \psi^{a|b}_{\alpha_j}\rangle|^2=p^{a}_{\alpha_j}\Leftrightarrow \psi^{a|b}_{\alpha_j}=\sqrt{p^{a}_{\alpha_j}}\left(\cos\gamma_{\alpha_j}+i\sin\gamma_{\alpha_j}\right)$ where $\gamma_{\alpha_j}=\arg{\psi^{a}_{\alpha_j}}, j\in\{1,2\}$
and this gives that
\begin{equation}\label{cosgamma}
\psi^{a|b}_{\alpha_2}\overline{\psi^{a|b}_{\alpha_1}}
=\sqrt{\paae \paat}\left(\cos\left(\gamma_{\alpha_2}-\gamma_{\alpha_1}\right)+i\sin\left(\gamma_{\alpha_2}-\gamma_{\alpha_1}\right)\right).
\end{equation}
The real part of the equations (\ref{Psipsi}) and (\ref{cosgamma}) gives
\begin{eqnarray}\label{ata}
\sqrt{\paae \paat}\cos&&\left(\gamma_{\alpha_2}-\gamma_{\alpha_1}\right)=\pabe\sqrt{\paaebe  \paatbe}-\pabt\sqrt{\paatbt\paaebt}
\\\nonumber
&-&\cos\theta_{\alpha_1}(\sqrt{\pabt\paatbt\pabe \paaebe}+\sqrt{\pabe \paatbe\pabt\paaebt}).
\end{eqnarray}
Moreover, since $\pabt=1-\pabe$ and from the condition that ${\bf P}^{a\vert b}$ is double stochastic i.e. $\paaebt=\paatbe=1-\paaebe=1-\paatbt$,  we rewrite (\ref{ata})
\begin{align}\label{nia}
\sqrt{\paae \paat}\cos\left(\gamma_{\alpha_2}-\gamma_{\alpha_1}\right)
&=\left(2\pabe-1\right)\sqrt{\paaebe(1-\paaebe)}\\\nonumber
&+\cos\theta_{\alpha_1}\left(1-2\paaebe\right)\sqrt{(1-\pabe)\pabe}.
\end{align}
Then by (\ref{GGTT}) and (\ref{GGTT1}) we obtain  $\cos\theta_{\beta_1}:$ 
\begin{equation}\label{tio2}
\cos\theta_{\beta_1}=\frac{\pbbe -{\pbae \pbbeae}-{\pbat\pbbeat}}{2\sqrt{\pbae \pbbeae \pbat\pbbeat}}.
\end{equation}
Multiply (\ref{tio2}) with $2\sqrt{\paae \paat}$ and use again that 
$\pabt=1-\pabe$ and ${\bf P}^{a\vert b}$ is double stochastic and 
\begin{equation}\label{elva2}
2\sqrt{\paae \paat}\cos\theta_{\beta_1}=\frac{\paae-1+\pabe+\pbaebe-2\pbaebe \paae}{\sqrt{\pbaebe \pbbeat}} .
\end{equation}
We will show that $\cos\left(\gamma_{\alpha_2}-\gamma_{\alpha_1}\right)=\cos\theta_{\beta_1}$	
or equivalent, we show that
 \begin{equation}
2\sqrt{\paae \paat}\cos\left(\gamma_{\alpha_2}-\gamma_{\alpha_1}\right)=2\sqrt{\paae\paat}\cos\theta_{\beta_1}.
\end{equation}
Multiply $\sqrt{\paae \paat}\cos\left(\gamma_{\alpha_2}-\gamma_{\alpha_1}\right)$ by $2\sqrt{\paaebe(1-\paaebe)}$ in the left-hand side (\ref{nia}) such that $LHS=2\sqrt{\paaebe(1-\paaebe)}$ $ \sqrt{\paae \paat} $ $\cos\left(\gamma_{\alpha_2}- \gamma_{\alpha_1}\right)$ and replace $\cos \theta_{\alpha_1}$ with $ \frac{\paae-{\pabe \paaebe}-{(1-\pabe)(1-\paaebe)}}{2\sqrt{\pabe \paaebe \pabt\paaebt}}$ in right-hand side
\begin{eqnarray}\label{yes}
LHS 
&=&2\left(2\pabe-1\right)\paaebe(1-\paaebe)\\\nonumber
&+&\left(\paae-{\pabe \paaebe}-{(1-\pabe)(1-\paaebe)}\right)\left(1-2\paaebe\right)\\\nonumber
&=&2\left(2\pabe-1\right)\paaebe(1-\paaebe)\\\nonumber
&+&\left(\paae-1+\pabe+\paaebe-2\pabe\paaebe\right)\left(1-2\paaebe\right)\\\nonumber
&=&2\left(2\pabe-1\right)\paaebe(1-\paaebe)\\\nonumber
&+&\left(\paae-1+\pabe+\paaebe-2\paaebe \paae\right)\\\nonumber
&-&2\paaebe\left(-1+\pabe+\paaebe-2\pabe\paaebe\right)-2\pabe\paaebe\\\nonumber
&=&2\left(2\pabe-1\right)\paaebe(1-\paaebe)\\\nonumber
&+&\left(\paae-1+\pabe+\paaebe-2\paaebe \paae\right)\\\nonumber
&-&2\paaebe\left(-1+2\pabe+\paaebe-2\pabe\paaebe\right)\\\nonumber
&=&\paae-1+\pabe+\paaebe-2\paaebe \paae.
\end{eqnarray}
From equation (\ref{elva2}) and (\ref{yes}) must.
\begin{eqnarray}
\frac{\paae-1+\pabe+\pbaebe-2\pbaebe\paae }{\sqrt{\pbaebe \pbbeat}}&=&\frac{\paae-1+\pabe+\paaebe-2\paaebe \paae}{\sqrt{\paaebe \pabeat}}\\\nonumber
&\Leftrightarrow&\\\nonumber
\pbaebe&=&\paaebe.
\end{eqnarray}
Therefore must $\cos\left(\gamma_{\alpha_2}-\gamma_{\alpha_1}\right)=\cos\theta_{\beta_1}$ iff $\mathbf{P}^{b|a}=\mathbf{P}^{a|b}$.
Let 
\begin{equation}\label{UAB}
U_{b \vert a}^{a \vert b}=\left(
\begin{array}{cc}
 \sqrt{\pbbeae} & \sqrt{\pbbeat} \\
 \sqrt{\pbbtae } & -\sqrt{\pbbtat}
\end{array}
\right).
\end{equation}
Then let us show that this vector is equivalent to the vector $\psi^{a \vert b}$ given by 
(\ref{EXYA}).
\begin{eqnarray} \label{UPSIi}
U_{b \vert a}^{a \vert b} \psi^{b \vert a}
&=& \sqrt{p^a_{\alpha_1}} e^{a\vert b}_{\alpha_1} +e^{i \theta_{\beta_1}} \sqrt{p^a_{\alpha_2}} e^{a \vert b}_{\alpha_2} \\\nonumber
&=& \sqrt{p^a_{\alpha_1}} e^{a\vert b}_{\alpha_1} +e^{i (\gamma_{\alpha_2}-\gamma_{\alpha_1})} \sqrt{p^a_{\alpha_2}} e^{a \vert b}_{\alpha_2}
\end{eqnarray} 
Then put $\psi^{a \vert b}_{\alpha_j}=\sqrt{p^a_{\alpha_j}}e^{i (\gamma_{\alpha_j})},j\in\{1,2\}$ into (\ref{EXYA})
\begin{eqnarray} 
\label{UPSIii}
\psi^{a \vert b}
&=& \sqrt{p^a_{\alpha_1}}e^{i (\gamma_{\alpha_1})} e^{a\vert b}_{\alpha_1} + \sqrt{p^a_{\alpha_2}} e^{i (\gamma_{\alpha_2})}e^{a \vert b}_{\alpha_2} \\\nonumber
&=& e^{i (\gamma_{\alpha_1})}U_{b \vert a}^{a \vert b} \psi^{b \vert a}
\end{eqnarray}
The complex amplitudes $\psi^{a \vert b}$ and $U_{b \vert a}^{a \vert b} \psi^{b \vert a}$ 
differs only by the multiplicative factor  $e^{i (\gamma_{\alpha_1})} \square$. Hence,
they belong to the same equivalent class of vectors on the unit sphere. Thus they are two representatives
of the same quantum state $\Psi^{b \vert a}.$
\begin{flushleft}
{\bf Acknowledgements}\\
I am grateful to my supervisor Professor Andrei Khrennikov for many 
discussions on the formulations of quantum mechanics and for introducing me into this field of research. 
I am also very thankful to Guillaume Adenier for all the discussions we have had
about the subject of quantum mechanics.
\end{flushleft}

\end{document}